\documentclass[twocolumn]{aastex631}
\usepackage{verbatim}
\usepackage{comment} 

\graphicspath{{./}{figures/}}

\def\gtsima{$\; \buildrel > \over \sim \;$}
\def\ltsima{$\; \buildrel < \over \sim \;$}
\def\gsim{\lower.5ex\hbox{\gtsima}} 
\def\lsim{\lower.5ex\hbox{\ltsima}}
\def\prosima{$\; \buildrel \propto \over \sim \;$} 
\def\simgt{\lower.5ex\hbox{\gtsima}} 
\def\simlt{\lower.5ex\hbox{\ltsima}} 
\def\simpr{\lower.5ex\hbox{\prosima}}
\def\msun{{\rm M}_{\odot}}

\revised{\today}

\submitjournal{ApJ}

\shorttitle{Black hole mass -- AGN luminosity scaling relations}
\shortauthors{Fiore et al.}

\begin{document}

\title{Constraining AGN accretion physics with black hole mass -- luminosity scaling relations}
\correspondingauthor{Fabrizio Fiore}
\email{fabrizio.fiore@inaf.it}

\author[0000-0002-4031-4157]{Fabrizio Fiore}
\affil{INAF-Osservatorio Astronomico di Trieste, via Tiepolo 11, I-34143 Trieste, Italy}
\affil{IFPU - Institute for Fundamental Physics of the Universe, via Beirut 2, I-34151 Trieste, Italy}

\author[0000-0003-2754-9258]{Massimo Gaspari}
\affil{Dip. di Fisica, Informatica e Matematica, Università di Modena \& Reggio Emilia, I-41125 Modena, Italy}

\author[0000-0002-2734-7835]{Simonetta Puccetti}
\affil{ASI – Agenzia Spaziale Italiana, Via del Politecnico snc, 00133, Roma, Italy}

\author[0000-0002-4314-021X]{Manuela Bischetti}
\affil{Dipartimento di Fisica "Enrico Fermi", Universita' di Pisa, Largo Bruno Pontecorvo 3, Pisa, I-56127, Italy}
\affil{INAF-Osservatorio Astronomico di Trieste, via Tiepolo 11, I-34143 Trieste, Italy}
\affil{IFPU - Institute for Fundamental Physics of the Universe, via Beirut 2, I-34151 Trieste, Italy}

\author[0000-0002-4227-6035]{Chiara Feruglio}
\affil{INAF-Osservatorio Astronomico di Trieste, via Tiepolo 11, I-34143 Trieste, Italy}
\affil{IFPU - Institute for Fundamental Physics of the Universe, via Beirut 2, I-34151 Trieste, Italy}

\author[0000-0001-9095-2782]{Enrico Piconcelli}
\affil{INAF-Osservatorio Astronomico di Roma, via Frascati 33, I00100, Monteporzio Catone, Italy}



\begin{abstract}
We test how supermassive black holes are fed by combining new black hole mass–luminosity relations with physically motivated feeding models. We build a uniform sample of 1,729 unobscured blue quasars ($z\!\lesssim\!2$) by cross-matching SDSS DR16 with eROSITA, and augment it with hyperluminous quasars (WISSH, HYPERION) plus 49 JWST broad-line AGN at $z>3.5$. We find for the SDSS-eRosita sample of blue quasars a near-linear scaling of bolometric luminosity with mass (slope $0.91\pm0.01$) and a shallower hard-X-ray trend (slope $0.73\pm0.01$). Classical hot-mode (Bondi) accretion underpredicts the observed luminosities by $\gtrsim2$ dex at the high-mass end and is inconsistent with the measured slopes. In contrast, Chaotic Cold Accretion (CCA)—in which multiphase gas condenses, collides, and rains onto the nucleus—consistently reproduces both the normalization and the near-linear slope expected from halo thermodynamics. The shallower X-ray relation points to a decreasing coronal power fraction with black hole mass. JWST broad-line AGN frequently appear X-ray weak or $H_{\alpha}$ enhanced. The latter case can be due to contributions from collisional ionization and photoionization from star-formation to the broad  $H_{\alpha}$ emission, leading to overestimate AGN luminosities and black hole masses. In the former case, the X-ray weakness is consistent with coronal shielding or anisotropy at high accretion rates. Overall, the data favor CCA-driven, self-regulated feeding over local spherical capture across $M_{\rm BH}\!\sim\!10^{7\text{–}10}M_\odot$, and motivate extending these tests to lower masses and higher redshifts.
\end{abstract}


\section{Introduction}
\label{sec:intro}

Active Galactic Nuclei (AGN) are among the most luminous objects in the universe, powered by the accretion of matter onto supermassive black holes (BHs) at the centers of galaxies. Accretion operates across multiple scales, from macroscales (kpc to Mpc, encompassing cluster or group halos), to mesoscales (pc to kpc, from the BH sphere of influence to galaxy scales), to microscales (from micro-parsec to parsec, from the BH horizon to the limit of the BH sphere of influence). The physics of such accretion processes on both macro-mesoscales and microscales, including angular momentum transport, feedback mechanisms via radiation pressure winds and jets, radiative efficiency, accretion rates, Eddington ratios, and BH spin remain a key area of research in astrophysics. 

Black hole mass-luminosity scaling relations link the BH mass ($M_{\rm BH}$) to the AGN luminosity (the bolometric luminosity $L_{\rm bol}$, or the luminosity in a given band, for example optical, UV, or X-ray), often through the Eddington luminosity ($L_{Edd} \propto M_{\rm BH}$), where the Eddington ratio $\lambda = L_{\rm bol} / L_{Edd}$ is the so-called Eddington ratio. It is remarkable that while both $L_{\rm bol}$ and $M_{\rm BH}$ span many orders of magnitudes, their ratio $\lambda$ has a relatively small range, as noted since the first reliable measurements of BH masses became available (see, e.g. \citealt{Wandel1999}). Black hole mass-luminosity scaling relations seem therefore a powerful tool to shed light on the physics of the accretion processes. 

Accretion onto SMBHs can proceed through two physically distinct channels operating across galaxy to halo scales (\citealt{Gaspari:2020}, for a review). In the \emph{hot mode}, diffuse X–ray emitting plasma (typically $T\sim10^{6}\text{--}10^{8}\,$K) drifts inward approximately quasi-spherically, where the angular momentum is low, a limit often modeled with the classical Bondi formalism \citep{Bondi:1952}. In this case, the characteristic capture radius is set by the sound speed, and the inflow rate is strongly dependent on the central gas entropy and on $M_{\rm BH}$, with additional suppression expected when rotation and turbulence are present. By contrast, the \emph{cold mode} arises when the hot halo precipitates into a multiphase medium – clouds and filaments with $T\sim10^{4}\,$K (and cooler) condense out of the ambient atmosphere and rain toward the nucleus, providing a low-angular-momentum fuel supply.

A leading framework for the cold channel is \emph{Chaotic Cold Accretion} (CCA), in which turbulent thermal instability in the galactic or group/cluster halo seeds recurrent condensation, cloud--cloud collisions promote rapid angular-momentum cancellation, and the net inflow becomes an ensemble process linked to the halo cooling rate rather than to local spherical capture \citep{Gaspari:2013_cca,Gaspari:2017_cca,Gaspari:2019,Wittor:2020}. CCA naturally predicts (i) intermittent, flickering feeding cycles; (ii) filamentary and clumpy cold structures coexisting with a hot atmosphere; and (iii) a scale-bridging connection between macro/meso fueling and micro-scale accretion. The prevalence of multiphase halos and filamentary cold gas in systems across mass and redshift, together with kinematic evidence for chaotic inflows, supports this precipitation-driven picture, as shown by a wide range of multiwavelength observations (e.g.; \citealt{Voit:2015_prof,Tremblay:2018,Maccagni:2021,Temi:2022,Wang:2023,Olivares:2022,Olivares:2025}). 

One of the most intriguing properties of CCA is its scale-invariance, where the accretion flow naturally proceeds from the mesoscales to the microscales. A similar fractal behavior has been suggested for the AGN nuclear wind outflow rate (normalized of the AGN accretion rate; \citealt{Fiore2024}). 

In contrast, in the standard scenario for AGN fueling, gas on galaxy scales is destabilized by processes such as galaxy interactions and/or internal galaxy dynamics (through e.g. bars and clumps), and is channeled toward galaxy nuclei (macro-meso scale accretion; see, e.g. \citealt{menci2014} and references therein). Once the gas enters an accretion disk, its further inward movement is primarily governed by viscous timescales, which is usually longer than the dynamic timescale. This effectively decouples macro from microscale accretion. Microscale accretion has been modeled at length using analytic approaches since the pioneering work of \cite{Shakura:1973}. Numerical approaches, including magnetohydrodynamic accretion flows, are more recent and span a relatively small parameter space (see e.g., \citealt{Davis:2020} for a review).  

In this paper, we first review both macro-, meso- and micro-scale accretion models and then compare their expectation to the BH-luminosity scaling relations built by using SDSS (DR16) and eROSITA surveys. The SDSS-eROSITA sample includes only unobscured blue quasars at redshift $<2$. We complement this sample with samples of hyperluminous quasars (the WISSH and HYPERION samples, \citealt{DegliAgosti2025,zappacosta23}) and with a sample of the broad line AGN at z$>3.5$ recently discovered by JWST. We use the bolometric/X-ray scaling relations $M_{\rm BH} - L_{\rm bol}$ and $M_{\rm BH} - L_{\rm x}$ to constrain both macro-mesoscale and microscale accretion models. We also use the $L_{\rm } - L_{H\alpha}$ scalings to study the apparent lack of X-ray emission in a large fraction of JWST $z>3$ AGN. 

\section{AGN samples}
\label{s:samples}
To study the AGN luminosity-BH mass scaling relations, we have considered the following AGN samples.

\subsection{SDSS-eROSITA sample}

We cross-correlated the SDSS DR16 quasar sample \citep{Wu:2022} with the eROSITA all sky survey catalog (eRASS1, \cite{Merloni:2024}), and with the eROSITA Final Equatorial-Depth Survey (eFEDS, \cite{Liu:2022}), using a separation $<12$arcsec (90\% of the matches have a separation $<9$arcsec). Considering eROSITA sources with detection maximum likelihood (DETML) $>12$ produces 12435 SDSS-eRASS1 matches and 2136 SDSS-eFEDS matches. We limited these samples to SDSS sources brighter than zmag=17.75 (for eRASS1 matches) and zmag=19 (for eFEDS matches). This ensures that $>80\%$ of the SDSS sources have an eRASS1 or eFEDS counterpart, minimizing issues related to low completeness. We further limited the samples to sources with E(B-V)$<0.05$ and $g-z<1$, to exclude obscured and/or red AGN. Finally, we limited the sample to sources with BH mass determinations obtained through single-epoch virial mass estimators employing the FWHM of the MgII and H$\beta$ lines, avoiding therefore BH mass determinations from the CIV line, which can be more uncertain because of a strong wind component in this line \citep{Coatman:2017}. The final samples include 1314 SDSS-eRASS1 AGN and 415 SDSS-eFEDS AGN, for a total of 1729 SDSS-eROSITA AGN at z=0.002-2. 

Finally, we compared this catalog with the catalogs of the optical counterparts of the eROSITA, eRASS1, and eFEDS sources of \cite{Salvato2025} and  \cite{Salvato2022}. The counterparts of the eFEDS sources in our catalog all match those in the \cite{Salvato2022} catalog. Regarding the SDSS counterparts of the eRASS1 sources, we found 1308 identical matches in the {\it Salvato\_etal2025\_DR1\_GDR3.colfits} catalog. The remaining six sources have a GDR3 counterpart, but 1 source has a different redshift and 5 sources lack spectroscopic identification in the {\it Salvato\_etal2025\_DR1\_GDR3.colfits catalog}. The separation between the eRASS1 sources and the SDSS quasars is between 1 and 6 arcsec, comparable with the separation distribution found for the full catalog. We visually inspected the SDSS spectra of these sources, which do not show peculiarities. We decided to keep the six sources in the sample, but we also verified that all the results presented in the paper do not depend on the inclusion or exclusion of these sources from our SDSS-eROSITA sample.

The 2-10 keV X-ray luminosity was calculated by converting the 0.2-2.3 keV fluxes to rest frame 2-10 keV fluxes for the eRASS1 sources and by converting the 0.5-2 keV luminosity to 2-10 keV luminosity for the eFEDS sources. A power law spectrum with an energy index $\alpha=1$ was used in the conversions for the results presented in the following sections. We verified that the results do not change assuming reasonable energy indices in the range 0.7-1.3. 

\subsection{WISSH sample}

The WISSH hyperluminous quasar sample was obtained by matching the WISE catalog with the SDSS catalog and selecting the 85 brightest sources at 22micron \citep{Saccheo:2023}. The WISSH sample includes quasars at z = 2-5 with bolometric luminosity $>10^{47}$erg/s. BH masses were evaluated from MgII and/or $H{\beta}$ lines observed in LBT/LUCI NIR spectrography (\cite{Vietri:2018}, Vietri priv.~comm.). X-ray luminosities are from \cite{zappacosta20} and \cite{DegliAgosti2025}.

\subsection{Hyperion sample}

The HYPERION sample includes highly accreting quasars at z$>6$ \cite{zappacosta23}. BH masses were evaluated from MgII lines. X-ray luminosities are from \cite{zappacosta20} and \cite{Tortosa2024}

\subsection {JWST sample}

We included in the JWST sample broad line AGN at $z>4$ from the JADES survey \citep{Maiolino:2024}, the FRESCO survey \citep{Matthee:2024}, and the UNCOVER survey (\cite{Greene:2024}, \cite{Bogdan:2024}. We also added to the sample the AGN from \cite{Juodzbalis:2024}, \cite{Napolitano:2024}, \cite{Larson:2023}, \cite{Li:2025}. 
BH masses and bolometric luminosities are estimated through the FWHM  and the luminosity of the H$\alpha$ line broad component using the relationships of \cite{Reines:2015} and \cite{Stern:2012}.

We used X-ray detection and the upper limits quoted in the cited papers. We also computed new upper limits for the UNCOVER sources in \cite{Greene:2024} and the JADES, CEERS and PRIMER sources in \cite{Li:2025}. There are a total of 49 AGN at z$>$3.5 in this sample, four of which at z$>8$. Only six of these AGN have an X-ray detection, see Table A1 in the Appendix. 

\section{Modeling the macro-meso scale feeding}
\label{s:model}

We describe here our main modeling and theories for the AGN bolometric luminosity versus BH mass scaling.

\subsection{Bondi (hot) accretion} 
\label{s:bondi}
Bondi theory is often used to describe the mode of accretion in AGN and is based on the quiescent/smooth spherically symmetric accretion of the central hot gas. The Bondi accretion rate $\dot M_{\rm B}$ is proportional to $\pi R_{\rm B}^2\rho v$ (with a normalization factor of order unity; \citealt{Gaspari:2013_cca}), where $\rho$ is the gas density and $v$ its gas radial velocity sufficiently far from the accretor (e.g. within the inner meso scale, $r < 1$ kpc). $R_{\rm B}$ is the Bondi radius, where the escape speed is equal to the sound speed\footnote{The sound speed is defined as $c_{\rm s} = (\gamma kT/\mu m_{\rm p})^{1/2}$, where $T$ is the gas temperature and $\mu\sim0.6$ the mean gas weight.}, thus setting the BH sphere of influence as $R_{\rm B}=2 G M_{\rm BH}/c_s^2$. For a non-relativistic gas adiabatic index of 5/3,\footnote{The Bondi pre-factor is 1/4, which cancels out the factor of 4 in the full Bondi formula.} the \citet{Bondi:1952} accretion rate is given by
\begin{equation}\label{eq:Md_Bondi}    
    \dot M_{\rm B} \simeq \pi(G M_{\rm BH})^2\,\rho/c_s^3,
\end{equation}
which is also proportional to the inverse of the gas entropy raised to the power of $-3/2$. To retrieve the BH mass scalings, as a normalization point, we choose the well-studied SMBH regime of $3\times10^9\ \msun$, which often resides in massive galaxies with mesoscale temperature and number density median values of 1\,keV and 0.1 cm$^{-3}$ (\citealt{Gaspari:2015_cca}); we add a typical (2$\sigma$) scatter of $\sim0.3$\,dex to both thermodynamical properties to account for mesoscale atmosphere diversity (\citealt{Gaspari:2019})\footnote{This study also shows that meso-scale entropy has no significant correlation with BH mass, unlike in the macro-scale halo case.}.

\subsection{Chaotic cold accretion} 
\label{s:cca}

One of the modern accretion theories emerging as consistent with several observables is Chaotic Cold Accretion (CCA -- \citealt{Gaspari:2013_cca,Gaspari:2020}, for a review).
Briefly, unlike classical hot-mode (Bondi, ADAF) or thin-disk (\citealt{Shakura:1973}) accretion models, CCA is based on the modern framework of complex systems. The diffuse halos of galaxies, groups, and clusters are expected to recurrently generate a precipitation of multiphase clouds and filaments via turbulent thermal instability (triggered by feedback and merger processes). This rain of chaotic clouds then increases the SMBH inflow rate via recursive inelastic collisions \citep{Gaspari:2015_cca,Gaspari:2017_cca}. 

Although CCA follows a complex detailed micro-evolution that requires high-resolution 3D hydrodynamical simulations, we can still capture the ensemble behavior via analytic prescriptions (\citealt{Gaspari:2019}) that are relevant for global scaling relations. As ensemble inflow\footnote{which is a superposition of multiple chaotic collisions}, CCA is directly proportional to the cooling rate of the macroscale ($0.1 R_{\rm vir}$\,$\sim$\,$100$\,kpc) diffuse gaseous halo (circumgalactic/intragroup core atmosphere) such as
\begin{equation}\label{eq:Md_CCA}    
    \dot M_{\rm cca} = q \frac{\mu m_{\rm p}}{(3/2)k}\,\frac{L_{\rm halo}}{T_{\rm halo}},
\end{equation}
where $q$ is a quenching factor that accounts for the flickering variability of the CCA, which simulations show to have median values of 0.2 with a scatter of 0.5 dex \citep{Gaspari:2017_cca}.  As in \S\ref{s:bondi}, we take a BH mass of $3\times10^{9}\ \msun$ as our normalization value, which corresponds to macroscale hot halo temperatures of $\sim$\,1 keV (\citealt{Gaspari:2019})\footnote{We note that the temperature is roughly isothermal for hot halos, hence its similarity with the meso-scale value.}. This also corresponds to the median halo luminosities of $6\times10^{42}$ erg\,s$^{-1}$, using the canonical halo scaling relations $L_{\rm halo} \approx 6\times10^{43} (T_{\rm halo}/2.2\,{\rm keV})^3$ erg\,s$^{-1}$ (\citealt{Sun:2012}). As the hot halo is dominated by the X-ray plasma emission, it is important not to confuse this \textit{macro} (100s kpc) halo X-ray luminosity with the \textit{micro} (sub-pc) AGN X-ray luminosity, which is usually much larger. To account for hot-halo diversity, we use a conservative 0.5 dex ($2\sigma$) scatter on the halo luminosity (\citealt{Gaspari:2019}).

In order to find the slope of our target scalings, we again leverage the tight hot-halo relations with SMBH mass by \citet{Gaspari:2019}, such that $M_{\rm BH} \propto T_{\rm halo}^{2.14}$. The final slope is thus given by $\dot M_{\rm cca} \propto {L_{\rm halo}}/{T_{\rm halo}} \propto T_{\rm halo}^2 \propto M_{\rm bh}^{0.94}$. As we will see in the Discussion \S\ref{s:disc}, such a nearly linear relationship  will be key to explaining the similar observational finding, at variance with the much steeper hot-mode predictions.

\subsection{Accretion rates to luminosities} \label{s:lum}
To convert the above accretion rates into our main observable -- luminosity -- we use the well-known relativistic rest-mass energy rate equation:
\begin{equation}\label{eq:L}
    L_{\rm bol} = \eta \dot M c^2,
\end{equation}
where $\eta$ is the radiative efficiency, $\dot M$ the given accretion rate (Bondi or CCA). 
To account for the diversity of radiative modes we use the canonical 0.1 value as median efficiency and add a wide 0.5 dex deviation to account for radiatively inefficient and rotating/Kerr modes at the lower and higher end, respectively (e.g.~\citealt{Sadowski:2016,Sadowski:2017}). 

In passing, we note that the final total scatter is obtained by combining each of the above scatters in quadrature and is shown in the main Fig.~\ref{lbollx} as shaded bands. Such conservative total scatter allows one to account for the large diversity in galactic/halo/accretion properties, without getting lost in the details of each theory/modeling, which purely act as additional scatter in the global scaling relation.

\section{Modeling micro-scale accretion}

CCA is scale invariant, meaning that microscale accretion is naturally connected with accretion on larger mesoscales and macroscales (statistically self-similar across scales). In contrast, in the classical scenario for AGN fueling, gas on galaxy scales is destabilized by processes such as galaxy interactions and/or internal galaxy dynamics and is channeled toward galaxy nuclei (macro-meso scale accretion). Once the gas enters an accretion disk, its further inward movement is primarily governed by viscous timescales, which is usually longer than the dynamical timescale. This effectively decouples macro from microscale accretion.  In the following, we review accretion disk models with the aim of providing a physical description of not only the observed bolometric luminosity but also of the X-ray luminosity. 

\subsection{Analytic accretion disk models}

Detailed accretion disk modeling with angular momentum transport (the so-called $\alpha$ disk model) was first provided by \cite{Shakura:1973}. A general-relativistic generalization soon followed thanks to \cite{Novikov:1973}, see for a review \cite{FKR:2002}. More recently \cite{Kubota:2018} presented optical to X-ray spectral energy distributions (SEDs) from accretion disks with \cite{Novikov:1973} emissivity, including both warm and hot Comptonization to account for AGN X-ray emission. \cite{Kubota:2019} extended the work to slim disks to account for super-Eddington AGN emission. In these models X-ray emission is produced by Comptonization of UV seed photons in a hot corona (see, e.g.~\cite{Haardt:1991}). The relative strength of the X-ray emission is a function of the corona size and temperature and of the fraction of accretion power dissipated in the corona. In the \cite{Kubota:2018} model this fraction is basically given by the relative extension of the inner corona with respect to the cooler disk.

In the $\alpha$-disk model, the bolometric luminosity is set by the accretion rate and the binding energy in the innermost stable orbit, so that $L_{\rm bol}\propto \dot M$ and, under the common assumption of a mass-independent Eddington ratio, $L_{\rm bol}$ scales roughly linearly with $M_{\rm BH}$.

These $\alpha$ disks are certainly a useful idealized tool to catch the main drivers of accretion around compact objects, but they are not realistic physical models, because of at least two main reasons: (1) realistic physical viscosity values are too small by many orders of magnitude with respect to that needed to explain the angular momentum transport; (2) the radiation-pressure dominated regions of disks are unstable due to thermal instabilities \cite{Shakura:1976}, leading to a limit-cycle behavior, which is not observed in AGN. For these reasons, numerical accretion disk models have been developed in the past twenty years, incorporating more physics, such as magneto-hydrodynamics (MHD).

\subsection{Numerical simulation of accretion disks}

Magnetorotational instabilities driving magnetohydrodynamic (MHD) turbulence together with MHD winds provide efficient angular momentum transport, see the review of \cite{Davis:2020}. One of the main results of MHD disk models is that magnetically supported disks do not develop thermal instabilities, as shown by \cite{Sadowski:2016} for accretion around 10 solar masses BHs and by \cite{Jiang:2019} for Eddington limited accretion on super-massive BHs. \cite{Jiang:2019b} extend this work to super-Eddington accretion ($\dot M= 250-520 L_{Edd}/c^2$). The \cite{Sadowski:2016} simulations use a general-relativistic (GR) MHD code. The \cite{Jiang:2019} and \cite{Jiang:2019b} simulations use a pseudo-Newtonian potential. More recently, \cite{Zhang:2025} presented full GRMHD simulations of radiation dominated accretion flows around stellar mass BHs.

\cite{Sadowski:2016} find that accretion disks where the magnetic field transport is not efficient, magnetic pressure, and radiation pressure cannot stabilize the disk, which quickly cools down and collapses towards the equatorial plane. In this case a hot corona is not formed, or in any case the fraction of accretion power dissipated in the corona is much smaller than that dissipated in the equatorial cooler disk. A hot corona is clearly formed in the \cite{Sadowski:2016} simulation where the magnetic field transport is efficient. The same happens in the \cite{Jiang:2019} and \cite{Jiang:2019b} simulations. In these models, the fraction of the accretion power dissipated in the hot, optically thin phase depends on the local disk surface density. If $k_R\Sigma/2<100$ (where $k_R$ is the Rosseland mean opacity and $\Sigma$ the local disk surface density), the fraction of accretion power dissipated in the hot phase can be large, even 50\% of the total, and a corona will form.

\cite{Jiang:2019b} and \cite{Jiang:2019} find that the hot corona is radially more extended at lower Eddington ratios. In particular, hot gas with temperature  $>10^7-10^8$K is present up to 20-25$R_{\rm g}$ for $\dot M/\dot M_{Edd}=0.07$, and up to $10-15R_{\rm g}$ $\dot M/\dot M_{Edd}=0.2$. At super-Eddington accretion rates, the corona is embedded in a funnel-like geometry, and from most viewing angles the corona is shielded by an optically thick gas. Furthermore, the corona is formed only in simulations in which angular momentum transport is provided by MRI turbulence, while in the case of accretion driven by spiral shocks the corona is faint or absent. \cite{Jiang:2019b} therefore predict that little X-ray emission can be observed from super-Eddington AGNs. In these simulations, the calculation of the X-ray spectrum and luminosity emitted from the corona is not carried out; therefore, it is not straightforward to compare the observed AGN X-ray luminosities with the expectations of these models. However, general consideration can be drawn, assuming that the X-ray luminosity is somewhat proportional to the corona size or that the X-ray spectrum depends on the corona geometry (see, e.g. \cite{Madau:2024}.

\cite{Zhang:2025} find that at near or sub-Eddington accretion in presence of net vertical magnetic flux the disk forms a thin, dense layer at midplane, surrounded by a magnetically dominated corona. In contrast, at super-Eddington accretion a geometrically thick, radiation-pressure supported disk forms, driving powerful outflows. The radiative efficiency of accretion drops from a few \% at sub-Eddington regimes to a fraction of \% at super-Eddington regimes. 

Numerical simulations so far were able to describe the accretion flow at micro-scales, but did not provide self-consistent broad band, optical to X-ray SEDs. For these reasons, we can only use these models to evaluate the coarse properties of the predicted emission: if the disk forms a dense, geometrically thin layer at midplane, it could produce a SED somewhat similar to that of analytic optically thick, geometrically thin disk models; if the disks form a hot corona, X-ray emission can be expected from this region, as Comptonization of cooler photons produced by the dense, thin layer is likely to occur. While we await for numerical models to produce quantitative radiation emission forecasts, we still have to resort to analytic models for more precise comparisons of the observed to model SEDs.

\begin{figure*}
\hspace{-0.7cm}
\includegraphics[width=9.5cm]{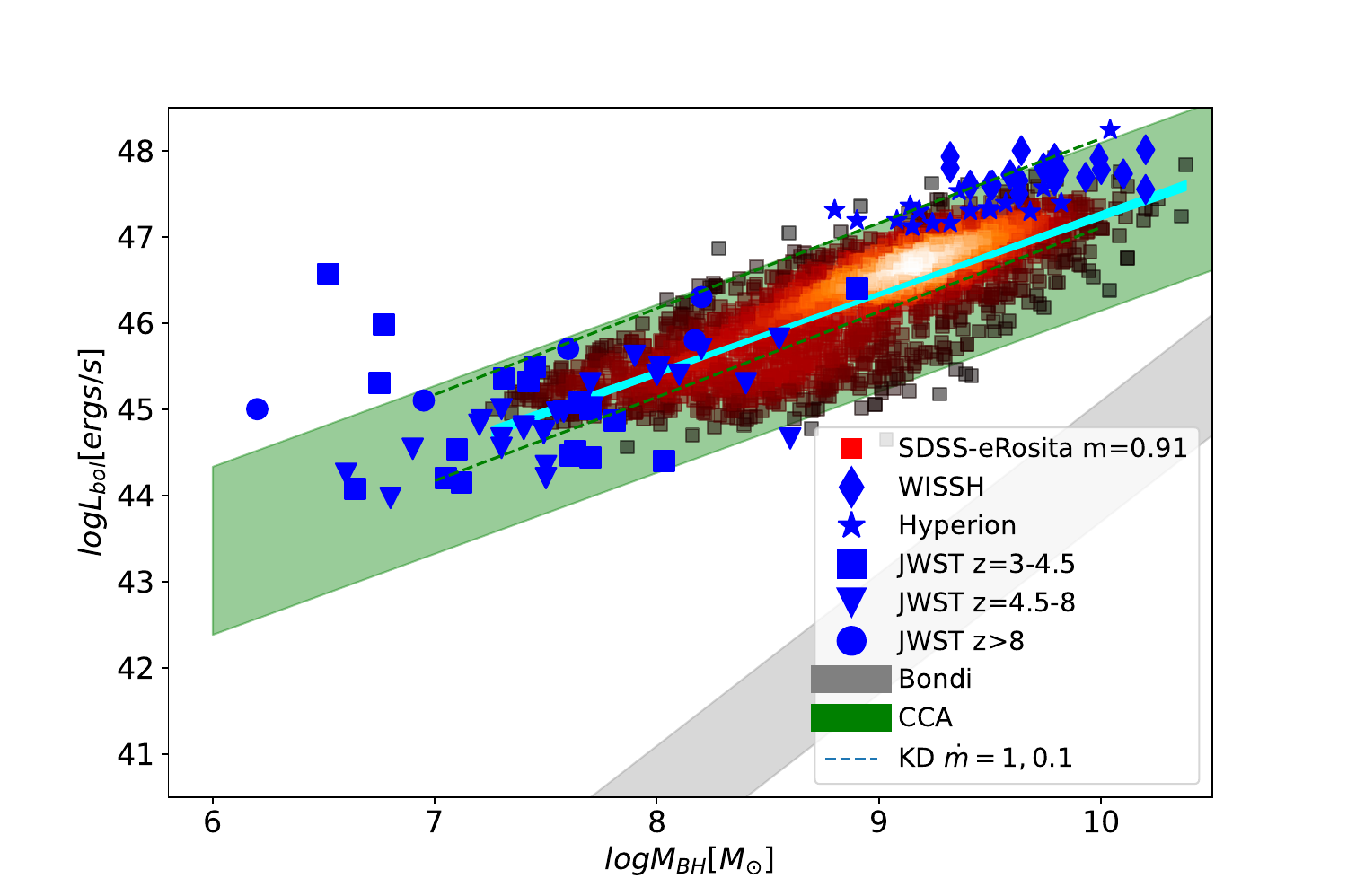}
\includegraphics[width=9.5cm]{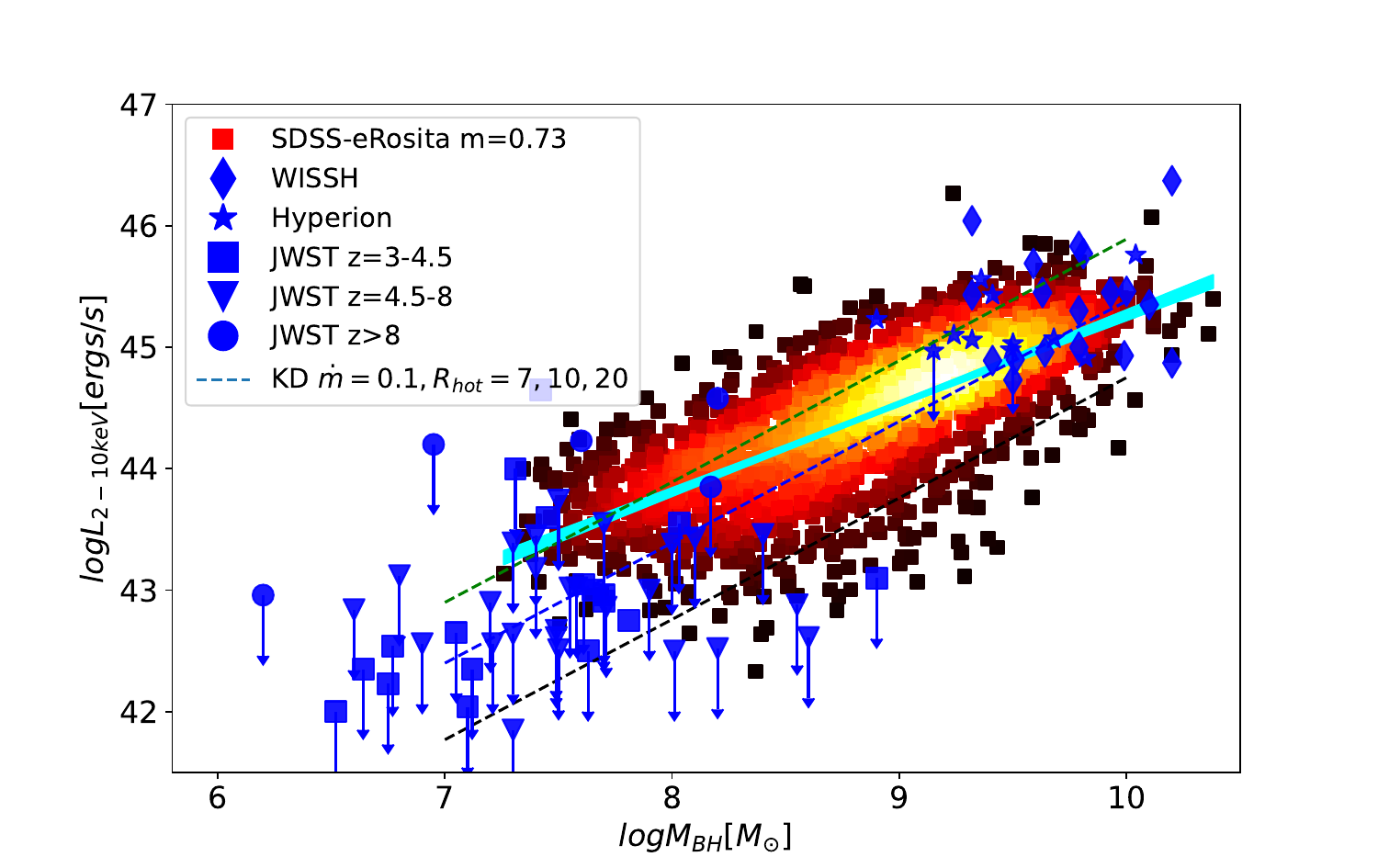}
 \caption{Bolometric luminosity ({\it left}) and X-ray (2-10 keV) luminosity ({\it right}) as a function of the BH mass for the AGN samples described in \ref{s:samples}. The cyan bow-ties represent the best fit regressions for the SDSS-eROSITA sample with slopes of 0.91$\pm$0.01 and 0.73$\pm$0.01 in the left and right panels respectively. The gray and green bands are the predictions of Bondi and CCA models, respectively (see \S\ref{s:disc}), with shaded band covering the $2\sigma$ (95\% level) scatter. The dashed lines are the expectation of the \citet{Kubota:2018} model for $\dot m=1.0,\,0.1$.} 
\label{lbollx}
\end{figure*}

\section {AGN black hole mass - luminosity scaling relations}
\label{s:scalings}

Figure \ref{lbollx} shows the bolometric and X-ray luminosities (2-10 keV), as a function of the BH mass for our AGN samples. The best-fit power-law slopes for the SDSS-eROSITA sample are 0.91$\pm0.01$ and 0.73$\pm0.01$ respectively (cyan bow tie in Fig. \ref{lbollx}). The scaling of $L_{\rm 2-10\,keV}$ with BH mass is significantly shallower than that of $L_{\rm bol}$, implying that the ratio of X-ray to bolometric luminosity of these blue quasars decreases with the BH mass. 

The prediction of the Bondi hot accretion mode is shown in Fig.~\ref{lbollx} as a gray band. The scaling of $logL_{bol}$ with $logM_{BH}$ is much steeper than what is observed, and the predicted bolometric luminosity is smaller than the observed one at all BH masses. Note that this refers to pure hot-mode radial accretion. If hot gas rotation and turbulence are considered, the Bondi mass accretion rate can be reduced by factors 3-4 \citep{Gaspari:2015_cca}.

In contrast, the prediction of CCA (green band) is consistent with the observations for normalization, slope of the $logL_{bol}-logM_{BH}$ scaling, and scatter around the scaling. 

The analytic accretion disk model of \cite{Kubota:2018} also predicts a scaling between $\log(L_{\rm bol})$ and $\log M_{\rm BH}$ close to that observed for the SDSS-eROSITA sample. The green dashed lines in Fig.~\ref{lbollx} refer to the scaling of a mass accretion rate of $\dot M=0.1-1$. 

The WISSH and HYPERION AGN are located at the upper end of the bolometric luminosity distribution of the SDSS-eROSITA AGN sample, reflecting their selection (the most luminous quasars at z=2-4 and the fastest accreting quasars at z$>$6, respectively).  In contrast, the JWST AGNs are broadly consistent with the $L_{\rm bol}$-$M_{\rm BH}$  distribution of the SDSS-eROSITA AGN sample. Note that the four JWST AGN at z$>8$ have $L_{\rm bol}$ at the upper end of the SDSS-eROSITA $L_{\rm bol}$ distribution. 

The X-ray emission is likely produced in a hot corona close to the BH and therefore only micro-scale accretion flow models can provide predictions. Numerical models can produce hot coronae in the innermost regions around BH, see previous section. However,  quantitative predictions of the expected X-ray luminosity and X-ray to bolometric luminosity ratio are not available. On the other hand, we can use predictions from analytic accretion disk models. Assuming a fixed corona size, the \cite{kubota18} model predicts a scaling of $\log L_{\rm x}$ with $\log M_{\rm BH}$ similar to the scaling of $logL_{bol}-logM_{BH}$ and thus steeper than observed. The observed $\log L_{\rm x}-\log M_{\rm BH}$ scaling might be recovered by assuming that $R_{\rm hot}$, the transition radius between the hot corona and the cooler disk decreases from $\sim$20 gravitational radii ($R_{\rm g}$) for BHs of $10^8 \msun$ to $\sim$10\,$R_{\rm g}$ for $10^{10} \msun$ BHs. Another possibility is that the innermost stable orbit of lower mass BHs is systematically smaller than that of higher mass BHs, or, in other words, that lower mass BHs spin at higher velocities than higher mass BHs. Intriguingly, \cite{King:2008} suggests that, on average, BH spin should decrease as the mass increases, because randomly oriented accretion episodes produce accretion disks co- or counter-aligned with the BH spin with similar frequencies, and, of course, the number of accretion episodes is higher for high BH masses.

The X-ray luminosity of the WISSH and HYPERION quasars is consistent with that of SDSS-eROSITA sources with BH mass similar to that of SDSS-eROSITA sources. However, among the 49 JWST sources there are only six X-ray detections, and many of the upper limits are below the area covered by the SDSS-eROSITA AGN, confirming the result of \cite{maiolino:2025}. In the next sections, we investigate these findings in more detail.

\section{Discussion}
\label{s:disc}

\subsection{Constraining the macro-meso scale accretion mode: Bondi vs.~CCA}

The gray band in Figure \ref{lbollx} is the expectation of Bondi accretion (\S\ref{s:bondi}). 
The expected luminosity (\S\ref{s:lum}) via the Bondi accretion rate at $M_{\rm BH}\sim10^{10}\,\msun$ is already 100$\times$ lower than the observed bolometric luminosity (left panel). This becomes substantially worse ($>$$1000\times$) toward lower BH masses and host galaxies, due to the steep Bondi scaling with $M_{\rm BH}^2$, which is ruled out by the observed relation, which has a slope of 0.91$\pm$0.01. More realistic gas dynamics, such as turbulence and rotation, would further suppress the effective Bondi rate by another 0.5-1 dex \citep{Gaspari:2015_cca}, thereby exacerbating the discrepancy. This does not necessarily imply that hot-mode accretion is absent but that over the full duty cycle of AGN, Bondi (`sunny weather') is expected to be highly subdominant, particularly in the intermediate to high-redshift range represented in our large samples to a significant extent. The comparison should not be interpreted as showing that hot gas is irrelevant. In the CCA picture, the hot halo remains the thermodynamic reservoir that regulates condensation. What is disfavored by the observed normalization and slope is instead classical, unboosted Bondi-like spherical capture from the local hot phase as the dominant fueling channel for this luminous blue-quasar population.

In contrast, the CCA predictions (\S\ref{s:cca}) are well consistent with the observed SDSS-eROSITA sample, from ultra-massive to intermediate-mass BHs (green band). This is due to two main properties. First, CCA can boost accretion rates by several dex over the Bondi/hot mode, since the feeding material is now the condensed dense cold gas coming from the macroscale halo, while Bondi is limited to very diffuse gas within the meso scale. Moreover, related inelastic collisions promote rapid accretion rate/luminosity boost with flickering variability (\citealt{Gaspari:2017_cca}), guaranteeing frequent observation of precipitation in large surveys.

Second, the slope generated by the CCA theory (via the cooling rate; \S\ref{s:cca}) over the full range of BH mass is quasi-linear (0.94), which preserves the sustained luminosities even for small SMBHs with masses up to $10^7\ \msun$. 
This near-linear slope arises from the thermodynamics of the hot halo, through the $L_{\rm halo}/T_{\rm halo}$ cooling–heating balance, rather than from local spherical capture, which is one of the key physics discriminators between CCA and Bondi accretion.
From a different perspective, this implies that the AGN feeding-feedback self-regulation is expected to be efficient (and self-similar) over a diverse range of systems (massive/elliptical galaxies to compact/spiral galaxies).

It is interesting to note that the driven CCA scatter ($\sim$0.9\,dex) is $\sim$25\% larger than that driven by Bondi accretion, mainly due to the additional flickering variability that is absent in hot-mode accretion. This increased scatter (green band), together with the diversity in the halo properties, represents well the overall (95\% level) dispersion of the eROSITA and other included samples. 


Overall, our novel observed scaling relations between either bolometric or hard X-ray luminosity correlated with BH mass rule out hot modes of accretion in most systems, while suggesting that CCA is a key driver of the AGN duty cycle over a wide range of galaxies, halos, and environments. This is in line with multiwavelength findings in other fields, which show that CCA is consistent with key observational probes, such as kinematical spectra and thermodynamical imaging of multiphase halos (e.g., \citealt{Tremblay:2018,Maccagni:2021,Temi:2022,Wang:2023,Olivares:2022,Olivares:2025}).

In the future, it will be important to extend such scaling relations to small-intermediate BH masses, to understand whether such quasi-linear scalings continue or not below the $M_{\rm bh} < 10^7\ \msun$ regime, i.e. whether precipitation still dominates over `sunny' (hot-mode) weather in large statistical samples of AGN.\\

\subsubsection{Implications for cosmological subgrid models}
Large–volume cosmological simulations typically adopt Bondi–Hoyle accretion often with highly ad-hoc “boosts” to mimic unresolved multiphase structure (e.g. \citealt{Booth:2009,Dubois:2014,Weinberger:2018}). While such boosts can tune the \emph{normalization} of BH growth, they retain the intrinsic $M_{\rm BH}^{2}$ (and entropy) dependence of hot-mode capture, which yields slopes that are too steep over $M_{\rm BH}\!\sim\!10^{7\text{–}10}\,M_\odot$ and overly smooth accretion histories. Variants that include angular-momentum suppression or torque-limited inflow improve the coupling to galaxy dynamics (e.g., \citealt{AnglesAlcazar:2015}), yet most frameworks still decouple BH feeding from halo thermodynamics. Our results argue for CCA-anchored prescriptions in which the ensemble inflow scales with the cooling budget of the hot atmosphere, e.g.~$\dot M_{\rm BH}\propto q\,L_{\rm halo}/T_{\rm halo}$ with $q$ drawn from a lognormal distribution to capture flickering (cloud–cloud collisions) and duty cycles \citep[e.g.,][]{Gaspari:2017_cca,Voit:2017}. Such CCA-linked models naturally recover the observed near-linear mass–luminosity slope and intrinsic variability without resorting to arbitrary boosts, providing a more faithful bridge between macro–meso fueling and emergent AGN observables.

\subsection{X-ray emission, $k_{\rm X}$, and Eddington ratios}

We found that the ratio of bolometric to X-ray luminosities increases with the BH mass for a large (1729 AGN) and well-selected sample of blue quasars at z$<2$ with reliable BH mass and bolometric luminosity estimates from SDSS spectroscopy and X-ray luminosity estimated homogeneously from eROSITA surveys. X-rays are thought to be produced by Compton scattering of seed UV photons in a hot corona. The observed $\log L_{\rm x}-\log M_{\rm BH}$ and $\log L_{\rm x}-\log L_{\rm bol}$ scalings can be recovered if the hot corona is smaller at high BH masses and/or if the fraction of accretion power dissipated in the corona is reduced at high BH masses.

\begin{figure*}
\includegraphics[width=9.5cm]{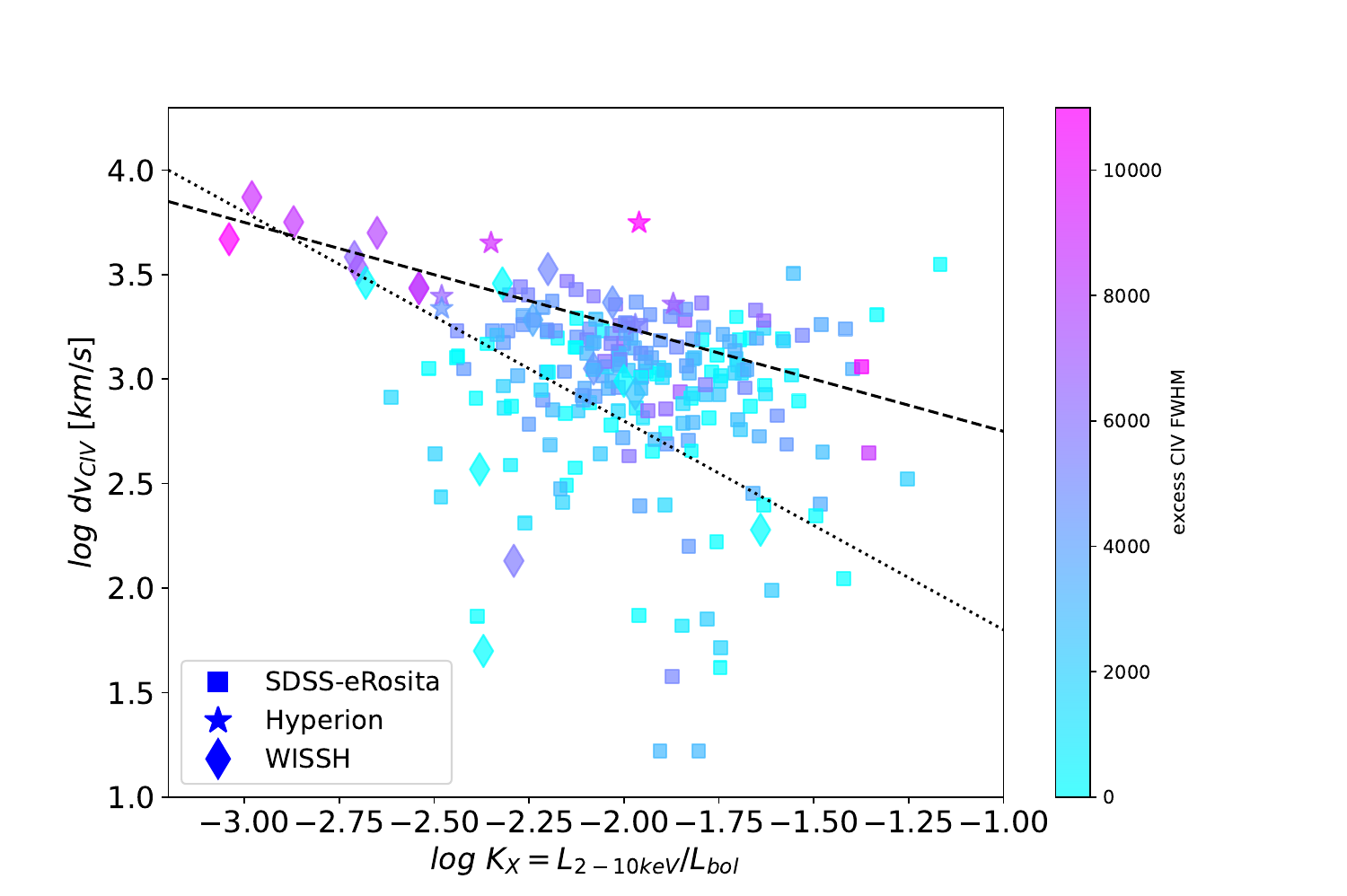}
\includegraphics[width=9.5cm]{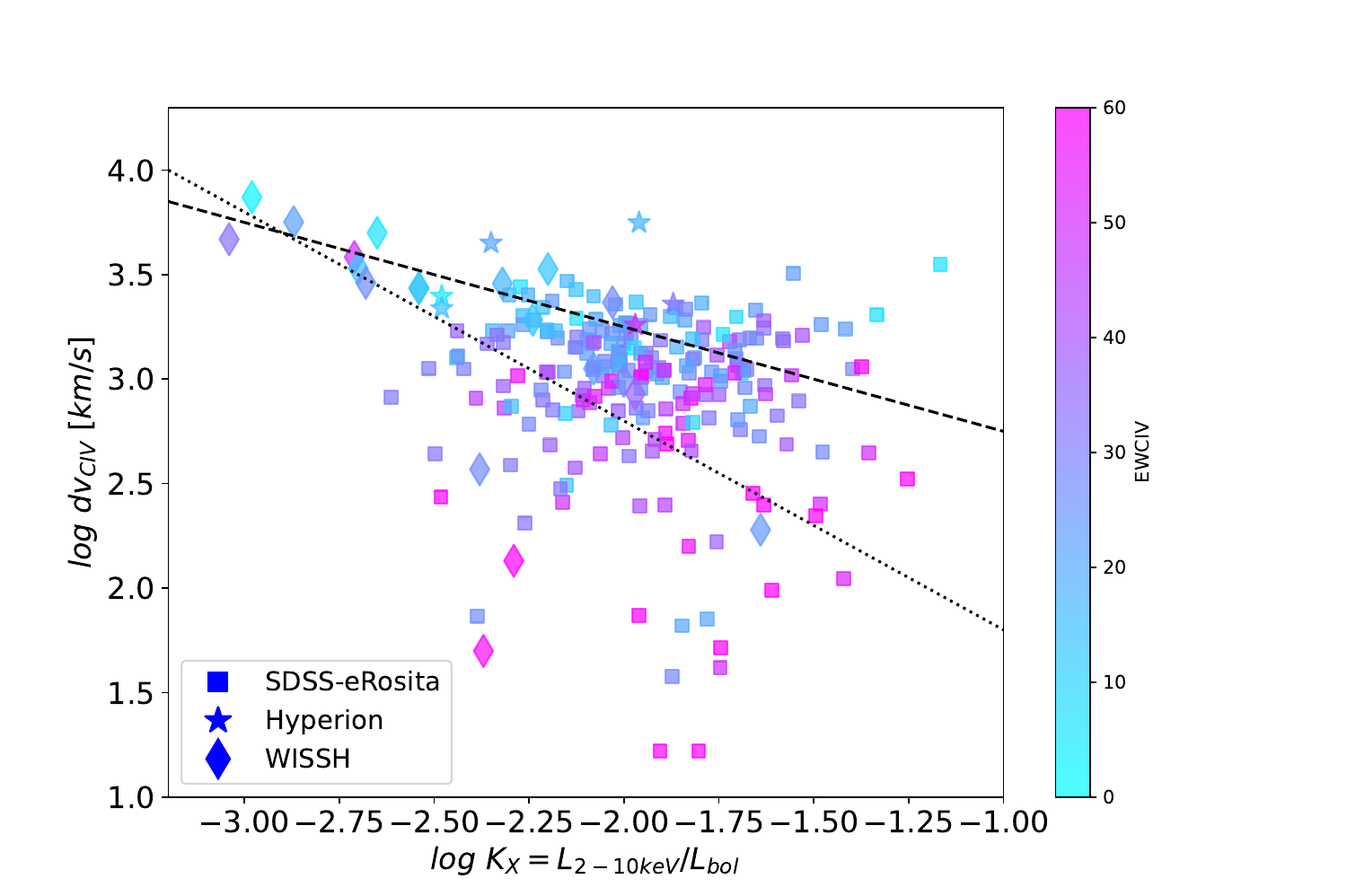}
\caption{The velocity shift between the CIV emission line centroid and the systemic redshift (measured from the MgII emission line centroid) as a function of $k_X=L_{2-10keV}/L_{bol}$ for a sample of 230 SDSS-eROSITA AGN (squares), WISSH quasars (diamonds) and HYPERION (stars) quasars. [Left panel]: color is proportional to the excess FWHM of the CIV line. The dashed and dotted lines have slopes -0.5 and -1 respectively. [Right panel]: color is proportional to CIV equivalent width }
\label{dvCIV}
\end{figure*}

A smaller corona can result from disturbances caused by nuclear winds and outflows, if the launching radii of these winds are comparable with the corona extension. For example, dense clumps in the winds can enhance the corona cooling via bremsstrahlung emission. Fast nuclear winds with column high densities, up to $10^{24}$ cm$^{-2}$ or even above this density, the so called Ultra Fast Outflows, have been observed in a large fraction of AGN and quasars (e.g. \cite{Tombesi:2010a}, \cite{Gianolli2024}, \cite{Fiore2024} and references therein). Recent XRISM observations showed that these winds can be highly clumpy, with several clouds along each line of sight \citep{xrism2025}. These dense clouds can enhance corona cooling and can efficiently shield further outflows from nuclear ionizing radiation, in particular X-rays, allowing for the presence of outflows that are not fully ionized, for example broad line region (BLR) winds, even at relatively small radii. In this case, a small/disturbed corona should be accompanied by UFOs and by fast BLR winds. A significant anticorrelation between the velocity shift of the CIV line centroid and the systemic redshift (a proxy for a BLR wind velocity) and the X-ray weakness (parameterized using $\alpha_{OX}$, the UV to X-ray power law index) has indeed been reported in the past (\cite{Vietri:2018}, \cite{zappacosta20}, \cite{DegliAgosti2025}. and references therein). We revisit this result in figure \ref{dvCIV}, showing the correlation between the velocity shift from CIV centroid and the centroid of a low ionization line (MgII), assumed to be close to systemic redshift, and $k_X=L_{2-10keV}/L_{bol}$ for 230 quasars in the SDSS-eROSITA sample with both CIV and MgII in the SDSS spectra and the WISSH and HYPERION samples. The color of the points in the left panel is proportional to the excess of CIV FWHM, calculated by subtracting in quadrature the MgII FWHM from the total CIV FWHM. If the MgII FWHM is dominated by virial gas motion, the excess CIV FWHM should thus be proportional to the BLR velocity, similarly to the CIV velocity shift. We see that the points on the upper envelope of the observed CIV shift - $k_X$, highlighted by the dashed line with slope -0.5, also have the high excess CIV FWHM. Points below this line tend to have a smaller excess FWHM, suggesting that the wide distribution of the CIV shift, in particular at medium to high $k_X$, may be due to inclination effects (with the sources seen at lower inclination angles $\theta$, showing the highest CIV shift and excess CIV FWHM, in line with previous determinations, see e.g. \cite{Bischetti:2017}, \cite{Vietri:2018}). This is also supported by the right panel of figure \ref{dvCIV}, where the color of the points is proportional to the equivalent width of the CIV emission line. Objects seen at larger inclinations should have higher CIV equivalent width, since the disk continuum is reduced by a factor $\approx cos(\theta)$,

\cite{Jiang:2019} using MHD global simulations of accretion flows around a $5\times 10^8\msun$ BH with zero spin find that the size of the corona depends on the accretion rate. However, they limit their analysis to a qualitative statement and do not compute X-ray emission from the corona. More work is clearly needed to investigate the parameter space further and to produce quantitative scaling predictions.

\begin{figure*}
\includegraphics[width=9.5cm]{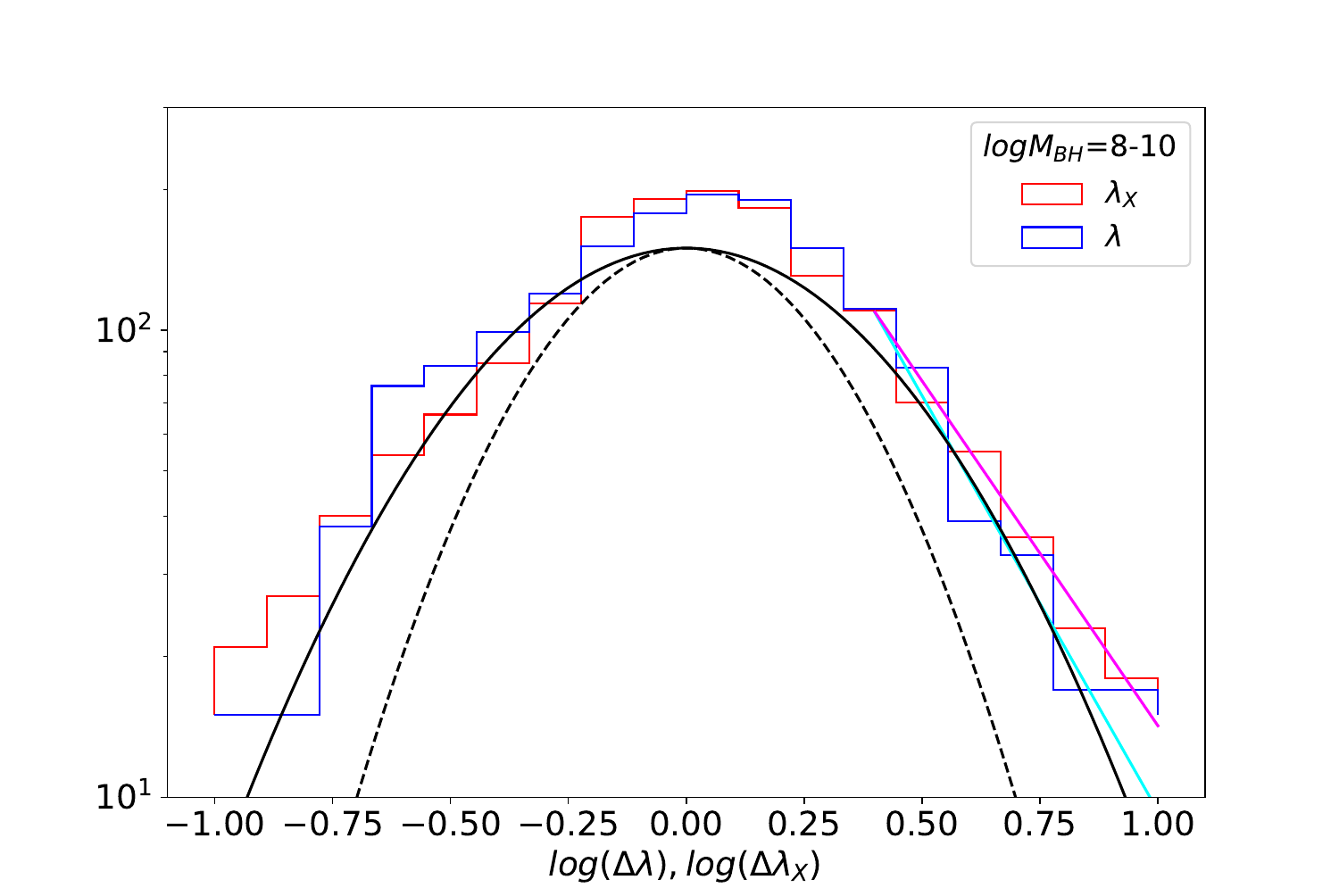}
\includegraphics[width=9.5cm]{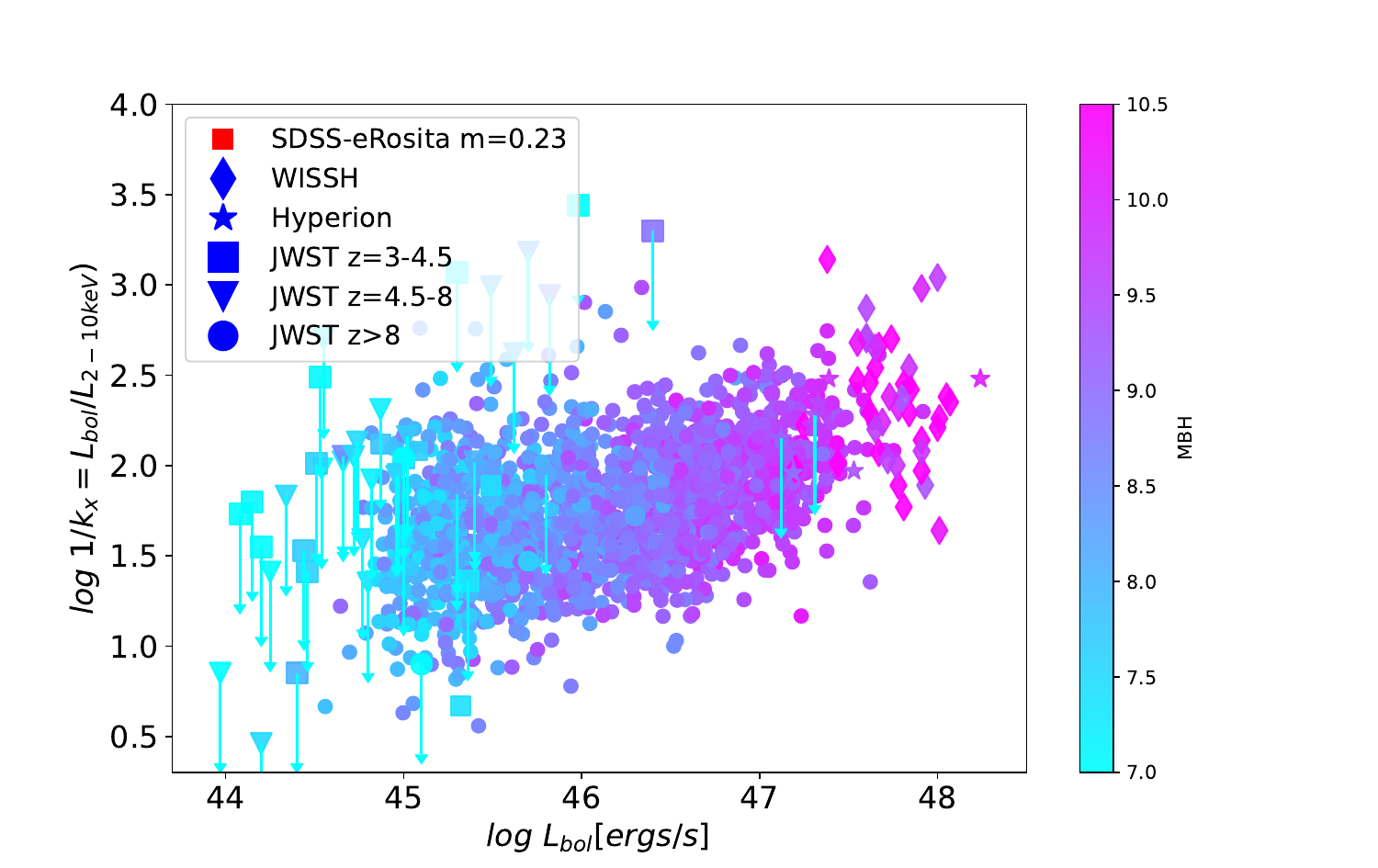}
\caption{[Left panel:] the distribution function of $\Delta \lambda$ (blue histogram) and $\Delta \lambda_X$ (red histogram). The cyan and magenta lines represent the best-fit power laws to the $\Delta \lambda$ and $\Delta \lambda_X$ tail distributions with slopes -1.8 and -1.5 respectively. The solid (dashed) black curve is a Gaussian function with $\sigma=0.4$ dex ($\sigma=0.3$ dex). 
[Right panel:] The correction between the X-ray luminosity (2-10 keV) and the bolometric luminosity for the AGN samples in figure \ref{lbollx}. Color of the symbols paints the BH mass (see right-hand colorbar).}
\label{lambdadistr}
\end{figure*}

We also found that the intrinsic scatter in the Eddington ratios of SDSS-eROSITA blue quasars, centered at $\lambda=L_{\rm bol}/L_{Edd}\sim0.1$ and $\lambda_X=L_{2-10keV}/L_{Edd}\sim0.03$ at $M_{\rm BH}=10^9 \msun$,  is relatively small, in any case comparable with the scatter in BH mass estimates. This is shown in figure \ref{lambdadistr}, left panel, showing the distribution of the deviations with respect to the best fit of the $\lambda-M_{\rm BH}$ and $\lambda_X-M_{\rm BH}$ relationships. The slopes of the distributions towards high $\lambda$ and $\lambda_X$ values are -1.8 and -1.5, respectively. The solid and dashed curves are Gaussian distributions with $\sigma=0.4$ dex and $\sigma=0.3$ dex, respectively, which bridges the typical scatter in single-epoch virial BH mass estimates, mainly due to the unknown system inclination. The scatter in the $\lambda$ and $\lambda_X$ distributions therefore appears to be dominated by the scatter in the BH mass estimates and any intrinsic scatter in the Eddington ratios is likely small.  Together, this implies that the variation of the X-ray to bolometric luminosity, that is, the $k_{\rm X}$ correction, must be dominated by the variations in BH mass and not by variation of the Eddington ratio; see figure \ref{lambdadistr}, right panel, showing the correlations between $1/k_{\rm x}$, $L_{\rm bol}$ and $M_{BH}$. We recall that the X-ray completeness of the SDSS-eROSITA blue quasar sample is about 80\%, with the remaining 20\% SDSS sources without eRosita detection likely having somewhat lower $L_X/L_{bol}$ values than that of the SDSS-eROSITA blue quasar sample. The true distributions of $\lambda_X$ and $k_{\rm X}$ will thus be somewhat broader than that in figure \ref{lambdadistr}. Furthermore, red quasars may have different  $\lambda_X$ and $k_{\rm X}$ distributions than the SDSS-eROSITA blue quasar sample.

High-redshift hyperluminous quasars (WISSH, HYPERION) are located at the upper border of the distribution covered by the SDSS-eROSITA AGN and touch/surpass the upper band of CCA in Figure \ref{lbollx}. In single-epoch virial BH mass estimates, if a constant "f" value is assumed for all inclinations, then objects observed at lower inclinations will have underestimated their BH mass, since the observed velocity dispersion is inclination dependent. An accretion disk seen nearly face-on will be brighter in UV than one seen at higher inclination angles. Since the adopted bolometric correction is an average one, the bolometric luminosity of face-on type-1 AGN may be overestimated \citep{Runnoe2012}. The actual Eddington ratios of the WISSH and HYPERION quasars should therefore be similar to the published one. These quasars may be tail events in the condensation rain variability, as they reach Eddington-level accretion, while low-z AGN and related average CCA rate tend to be in the sub-Eddington regime.

\subsection{JWST broad line AGN at high redshift}

\begin{figure}
\includegraphics[width=9.5cm]{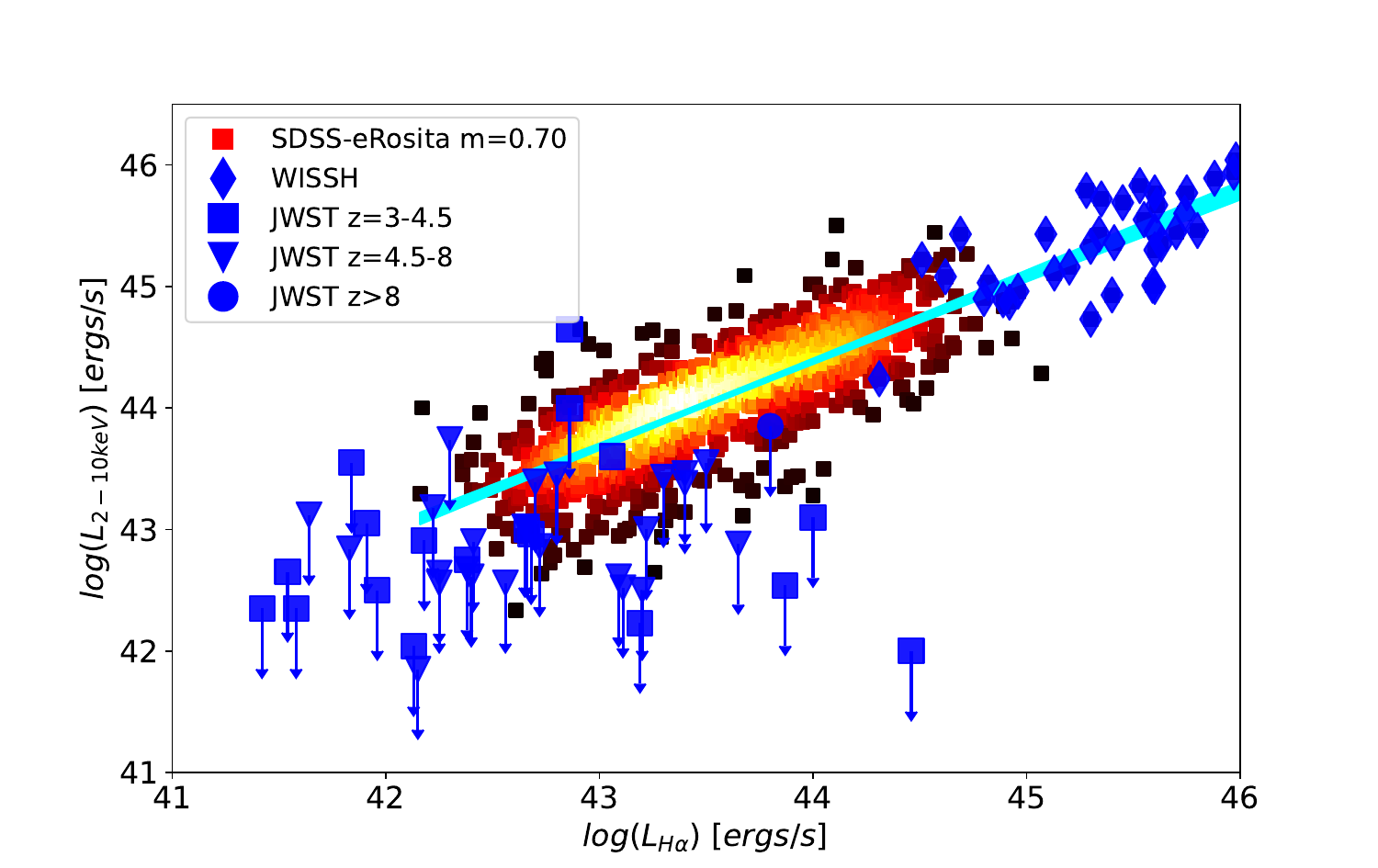}
\caption{The X-ray 2-10 keV luminosity as a function of the broad H$\alpha$ line for a sample of 945 SDSS-eROSITA AGN (red-yellow squares), WISSH quasars (blue diamonds) and JWST AGN at z$>$4 (blue squares, circles and triangles). Arrows denote upper limits. For about half of the SDSS-eROSITA AGN the H$\alpha$ line was in the range of the SDSS spectrum, for the remaining objects we scaled the luminosity of the $H\beta$ line for the average $H\alpha/H_\beta$ ratio of 3.55 found for the quasars with both H$\alpha$ and $H_\beta$ in the SDSS spectra.}
\label{lxlha}
\end{figure}

JWST has recently found a population of broad-line AGNs with redshifts up to z = 10 (\cite{maiolino:2025} and references therein). We have collected a sample of 49 broad line AGNs at $z>3.5$ (4 AGNs at z$>8$). We used X-ray luminosities and upper limits from the literature for 29 AGNs and calculated luminosities and upper limits for 9 UNCOVER sources and 11 sources in \cite{Li:2025}. The bolometric luminosity of all JWST AGN was calculated by scaling the luminosity of the broad components of the H$\alpha$ line; therefore, it can be instructive to study the correlation between the observed quantities. Figure \ref{lxlha} shows the correlation between the X-ray luminosity and the luminosity of the broad component of the H$\alpha$ line for the JWST broad line AGNs and for the 945 SDSS-eROSITA AGN. For about half of the latter AGN the H$\alpha$ line was in the range of the SDSS spectrum, for the remaining objects we scaled the luminosity of the $H\beta$ line for the average $H\alpha/H_\beta$ ratio of 3.55 found for the quasars with both H$\alpha$ and $H_\beta$ in the SDSS spectra. We added to this sample the WISSH quasars with NIR spectroscopy covering either H$\alpha$ or $H_\beta$ transitions. We find $\log L(2-10{\rm keV})\propto 0.7\,\log L({\rm H\alpha})$, similar to the scaling found in \cite{Jin:2012}. Many JWST sources are below the best-fit correlation, which implies that they are faint X-ray sources, broad H$\alpha$ enhanced sources, or both (also see \cite{Tortosa2026}). Broad H$\alpha$ AGN emission is usually associated with photo-excitation of gas clouds with density $>10^8$cm$^{-3}$. However, at high densities and high temperatures, or if the emitting gas is shielded from the ionizing continuum, collisional excitation can be important and can contribute to the observed line emission. Furthermore, the host galaxies of JWST high-redshift AGN are usually very compact, and therefore the broad emission line clouds may be also excited by star-formation continuum. The star-formation rate in the host galaxies of the JWST AGN cannot be measured easily, since it requires a detailed comparison of the observed UV-to-NIR spectral energy distribution with galaxy and AGN templates. \cite{Xiao:2024} provide star-formation rates for a sample of 7 FRESCO AGNs considered in our sample, with values between 1.5 and 23 $M_{\odot}$\,yr$^{-1}$, corresponding to a UV luminosity in the range $10^{43}-10^{44}$\,ergs\,s$^{-1}$, comparable to the observed UV luminosity. The star-formation contribution to broad H$\alpha$ excitation may then be relevant, at least in these sources. 

In summary, some of the JWST AGN may have Balmer lines enhanced by collisional ionization and/or photoionization from star-formation. If this is the case, their bolometric luminosity may have been overestimated (by even one or two orders of magnitude), and they may not be intrinsically weak X-ray sources but simply lower luminosity AGN. Also, their black hole mass was estimated using the broad H$\alpha$ luminosity and H$\alpha$ FWHM. More in detail, according to \cite{Reines:2015} the black hole mass should scale with about the square root of the broad H$\alpha$ luminosity and the square of the H$\alpha$ FWHM. Therefore, if the broad H$\alpha$ luminosity is enhanced with respect to simple AGN photoionization models, also the black hole masses are overestimated. Furthermore, electron scattering can also be efficient in broadening an intrinsically narrow  H$\alpha$ line at high electron densities \citep{Rusakov2026}.

On the other hand, as discussed in previous sections, the X-ray luminosity depends on the size and geometry of the hot corona.  In particular, \cite{Jiang:2019b} show that at high accretion rates the corona is embedded in a funnel-like geometry and that it is shielded by an optically thick gas from most viewing angles, strongly reducing the X-ray emission observed from most lines of sight. In any case for such corona geometry the X-ray emission would have a steep spectral index \citep{Madau:2024}, further reducing the observed X-ray luminosity in the 2-10 keV band.  Furthermore, faint X-ray emission is also expected at sub-Eddington rates for particular configurations of the input magnetic field \citep{sadowski16}. 

\section{Conclusions}

Using a uniform set of 1,729 SDSS–eROSITA blue quasars ($z\!\lesssim\!2$), augmented by hyperluminous quasars and 49 JWST broad-line AGN at $z>3.5$, we establish two tight mass–luminosity relations: a nearly linear $L_{\rm bol}$–$M_{\rm BH}$ (slope $0.91\pm0.01$) and a shallower $L_{2-10}$–$M_{\rm BH}$ (slope $0.73\pm0.01$). Because of the strong correlation between redshift, bolometric and X-ray luminosities in our bright quasar sample, a similar strong correlation is present between redshift and BH masses. Thus, we are not able to investigate possible evolutions of the luminosity - BH mass correlations with the redshift with the current quasar sample.

The observed dispersions are small and consistent with single-epoch virial systematics, implying limited intrinsic scatter, characteristic Eddington ratios of order $\sim0.1$, and, as a consequence, an X-ray bolometric correction that increases with $M_{\rm BH}$.

Hot-mode (Bondi-like) capture cannot account for these trends: it underpredicts $L_{\rm bol}$ by over 2 orders of magnitude at the high-mass end and, more importantly, imposes a quadratic mass dependence ($\propto M_{\rm BH}^{2}$) that is at odds with the measured slopes, even before including the further suppression expected from rotation and turbulence. In contrast, a feeding cycle driven via CCA rain links the ensemble inflow to halo thermodynamics ($\dot{M}\!\propto\!L_{\rm halo}/T_{\rm halo}$), which naturally yields a \emph{near-linear} mass scaling and the observed level of variability/flickering. The core result is therefore that macro/meso halo physics (not local spherical capture) sets the mass scaling of the AGN power.

The shallower $L_{2-10}$ relation indicates a declining hard-X fraction with increasing mass, consistent with a smaller coronal lever arm, possibly due to disturbances from nuclear winds and outflows, and/or systematic trends in spin or magnetic topology at high $M_{\rm BH}$. 

At high redshift, the X-ray luminosity of many JWST broad-line AGN appears much lower than that expected from the bolometric luminosity estimated by scaling the luminosity of the broad component of the $H_{\alpha}$ line. However, this luminosity can be boosted by both collisional ionization and photoionization from star-formation. If this is the case, these AGNs may simply be lower luminosity AGNs. In cases where X-ray weakness is present, this could be explained by high accretion-rate flows adopting funnel-like geometries that self-shield the corona and producing intrinsically softer spectra.

Taken together, the data favor CCA-regulated feeding across $M_{\rm BH}\!\sim\!10^{7\text{--}10}\,M_\odot$. Next steps may include (i) pushing to lower masses and higher redshifts, (ii) joint measurements of host-halo and AGN properties, and (iii) securing coronal diagnostics ($E_{\rm cut}$, reverberation, anisotropy) to pin down the origin of the hard-X flattening. These benchmarks will also inform subgrid choices in cosmological simulations, helping to assess whether precipitation–driven accretion, which links fueling to halo thermodynamics, provides a more realistic framework than ad-hoc modified Bondi prescriptions.

\vspace{4mm}
{\bf Acknowledgments}

This work was performed in part at Aspen Center for Physics, which is supported by National Science Foundation grant PHY-2210452. FF thanks Chris Done and Guido Risaliti for useful discussions. FF, CF \& MB acknowledge support from PRIN MUR 2022 2022TKPB2P - BIG-z. M.G. acknowledges support from the ERC Consolidator Grant \textit{BlackHoleWeather} (101086804). EP acknowledges support from the Large Program
“DELUX” of the “Ricerca Fondamentale 2024” INAF program.

This work uses data from eROSITA, the soft X-ray instrument aboard SRG, a joint Russian-German science mission supported by the Russian Space Agency (Roskosmos), in the interests of the Russian Academy of Sciences represented by its Space Research Institute (IKI), and the Deutsches Zentrum für Luft- und Raumfahrt (DLR). The SRG spacecraft was built by Lavochkin Association (NPOL) and its subcontractors, and is operated by NPOL with support from the Max Planck Institute for Extraterrestrial Physics (MPE). The development and construction of the eROSITA X-ray instrument was led by MPE, with contributions from the Dr. Karl Remeis Observatory Bamberg \& ECAP (FAU Erlangen-Nuernberg), the University of Hamburg Observatory, the Leibniz Institute for Astrophysics Potsdam (AIP), and the Institute for Astronomy and Astrophysics of the University of Tübingen, with the support of DLR and the Max Planck Society. The Argelander Institute for Astronomy of the University of Bonn and the Ludwig Maximilians Universität Munich also participated in the science preparation for eROSITA. 

This work uses SDSS catalogs. Funding for the Sloan Digital Sky Survey V has been provided by the Alfred P. Sloan Foundation, the Heising-Simons Foundation, the National Science Foundation, and the Participating Institutions. SDSS acknowledges support and resources from the Center for High-Performance Computing at the University of Utah. SDSS telescopes are located at Apache Point Observatory, funded by the Astrophysical Research Consortium and operated by New Mexico State University, and at Las Campanas Observatory, operated by the Carnegie Institution for Science. The SDSS website is www.sdss.org.

SDSS is managed by the Astrophysical Research Consortium
for the Participating Institutions of the SDSS Collaboration,
including Caltech, The Carnegie Institution for Science,
Chilean National Time Allocation Committee (CNTAC) ratified
researchers, The Flatiron Institute, the Gotham Participation
Group, Harvard University, Heidelberg University, The
Johns Hopkins University, L’Ecole polytechnique federale de
Lausanne (EPFL), Leibniz-Institut fur Astrophysik Potsdam
(AIP), Max-Planck-Institut fur Astronomie (MPIA Heidelberg),
Max-Planck-Institut fur Extraterrestrische Physik (MPE), Nanjing University, National Astronomical Observatories of China (NAOC), New Mexico State University, The Ohio State University, Pennsylvania State University, Smithsonian Astrophysical Observatory, Space Telescope Science Institute (STScI), the Stellar Astrophysics Participation Group, Universidad Nacional Autonoma de Mexico, University of Arizona, University of Colorado Boulder, University of Illinois at Urbana-Champaign, University of Toronto, University of Utah, University of Virginia, Yale University, and Yunnan University.

\vspace{5mm}
\facilities{eROSITA, SDSS, JWST, Chandra}

\software{astropy \citep{2013A&A...558A..33A,2018AJ....156..123A},  
TOPCAT \url{http://www.starlink.ac.uk/topcat/}, CIAO, CALDB 
          }

\bibliography{bibliography,biblio}

\appendix
\setcounter{table}{0}
\renewcommand{\thetable}{A\arabic{table}}

\section{Appendix}

To evaluate X-ray fluxes and upper limits for the UNCOVER sources \citep{Greene:2024} and the JADES, CEERS, and PRIMER sources from \cite{Li:2025}, we retrieved data for the CDF-S, CDF-N, UDS, and Abell 2744 fields from the Chandra Archive (https://cda.cfa.harvard.edu/chaser/). These datasets comprise 102, 20, 25, and 104 observations, respectively. Data processing was performed using the Chandra Interactive Analysis of Observations (CIAO) software (v4.17.0) and CALDB (v4.11.6), following the official analysis procedures provided by the Chandra X-ray Center (https://cxc.cfa.harvard.edu/ciao/threads/index.html).

We evaluated the source count rate in the 0.5-2 keV energy band using aperture photometry, with a source aperture radius of at least 2\". We applied corrections for the lost encircled count fractions using the PSF appropriate to each observation. For sources with a signal-to-noise ratio less than 3, we derived a 90\% upper limit following the procedure described in \cite{Narsky2000}.

Count rates in the 0.5–2 keV band were converted to 2–10 keV rest-frame count rates using a power-law spectral model with a photon indices $\Gamma = 1.4$ and $\Gamma=2.3$. We used for the analysis presented in this paper the average luminosities (see Table \ref{XWebb}) and checked that the uncertainty on this value due to the unknown photon index is $<0.1-0.3$ dex (depending on the redshift). 

Table \ref{XWebb} gives for the 49 JWST AGN the redshift, the X-ray 2-10 keV and bolometric luminosities and the black hole masses estimated from the broad $H_{\alpha}$ line luminosities and FWHM.

\begin{table}[]
    \centering
    \begin{tabular}{llllll}
\hline
Name & redshift & $logL_{2-10keV}$ & $logL_{bol}$ & $logM_{BH}$ & note\\
\hline
GS10013704	& 5.919	& $<$42.51 & 44.34 & 7.5 & 1,2 \\
GS8083	& 4.647	& $<$41.85 & 44.55 & 7.3 & 1,2\\
GN1093	& 5.594	& $<$43.18 & 44.77 & 7.4 & 1,2\\
GN3608	& 5.269	& $<$43.12 & 43.97 & 6.8 & 1,2\\
GN11836	& 4.409	& $<$42.04 & 44.53 & 7.1 & 1,2\\
GN20621	& 4.682	& $<$42.64 & 44.66 & 7.3 & 1,2\\
GN73488 & 4.133	& $<$42.96 & 45.0 & 7.7 & 1,2\\
GN77652	& 5.229	& $<$42.56 & 44.54 & 6.9 & 1,2\\
GN61888 & 5.874 & $<$42.90 & 44.82 & 7.2 & 1,2\\
GN62309	& 5.151	& $<$42.84 & 44.25 & 6.6 & 1,2\\
GN53757	& 4.447 & $<$42.91 & 44.44 & 7.7 & 1,2\\
GN954	& 6.759 & $<$43.00 & 45.62 & 7.9 & 1,2\\
GS3073	& 5.55	& $<$42.52 & 45.70 & 8.2 & 2\\
GN-z11	& 10.604 & $<$42.96 & 45.0 & 6.2 & 3\\
GN1001830 &	6.68 & $<$42.61 &	44.66 &	8.6	& 4\\
GS17341 & 3.598 & $<$42.0  & 46.57 & 6.5 & 2\\
GN2916  & 3.664	& $<$42.54 & 45.98 & 6.8 & 2\\
GS13329	& 3.936 & $<$42.23 & 45.30 & 6.7 & 2\\
GN4014	& 5.228	& $<$43.02 & 44.97 & 7.6 & 5\\
GN9771	& 5.538	& $<$42.88 & 45.82 & 8.5 & 5\\
GN12839	& 5.241 & $<$42.5	 & 45.49 & 8.0 & 5\\
GN13733	& 5.236	& $<$42.67 & 44.72 & 7.5 & 5\\
GN14409	& 5.139	& $<$42.56 & 44.87 & 7.2 & 5\\
GN15498	& 5.086 & $<$42.86 & 45.02 & 7.7 & 5\\
GN16813	& 5.355	& $<$43.02 & 44.96 & 7.5 & 5\\
GS13971	& 5.481	& $<$42.61 & 44.74 & 7.5 & 5\\
UHZ1	& 10.071 & 44.23 & 45.7  & 7.6 & 6\\
GHZ9	& 10.145 & 44.58 & 46.30 & 8.2 & 7\\
Ceers1019 & 8.68 & $<$44.2 & 45.1 & 6.9 & 7\\
U4286	& 5.84  & $<$43.38 & 45.4 & 8.0 & 8,9\\
U13123	& 7.04	& $<$43.39 & 45.0	& 7.3 & 8,9 \\
U13821	& 6.34	& $<$43.43 & 45.4 & 8.1 & 8,9\\
U20466	& 8.502	& $<$43.85 & 45.8 & 8.2 & 8,9\\
U23608	& 5.8	& $<$43.74 & 44.2 & 7.5	 & 8,9\\
U35488	& 6.26	& $<$43.45 & 44.8 & 7.4 & 8,9\\
U38108	& 4.96	& $<$43.46 & 45.3 & 8.4	& 8,9\\
U41225	& 6.76	& $<$43.55 & 45.3 & 7.7	& 8,9 \\
U45924	& 4.46	& $<$43.1 & 46.4 & 	8.9	 & 8,9 \\
13025-JADES-GS	& 3.472 & $<$42.65 & 44.20 & 7.0 & 10,9 \\
27578-JADES-GS	& 3.704 & 42.75 & 44.87 & 7.8 & 10,9\\
24748-JADES-GS	& 3.704 & 43.60 & 45.49 & 7.4 & 10,9\\
70911-JADES-GS	& 3.672 & 43.0 & 45.08 & 7.6 & 10,9\\
24735-JADES-GS	& 3.584 & $<$42.35 & 44.15 & 7.1 & 10,9\\
32092-JADES-GS	& 3.473 & $<$42.35 & 44.08 & 6.6 & 10,9\\
63546-JADES-GS	& 3.605 & $<$42.50 & 44.51 & 7.6 & 10,9\\
7115-JADES-GN	& 4.128 & $<$43.05 & 44.46 & 7.6 & 10,9\\
19400-PRI-UDS	& 3.698 & $<$43.55 & 44.40 & 8.0 & 10,9\\
7435-PRI-UDS	& 3.982 & $<$44.0 & 45.36 & 7.3 & 10,9\\
81571-PRI-UDS	& 3.942 & 44.65 & 45.32 & 7.4 & 10,9\\
\hline
    \end{tabular}
    \caption{1=\cite{Maiolino:2024}; 2=\cite{maiolino:2025}; 3=\cite{maiolino24nat}; 4=\cite{Juodzbalis:2024}; 5=\cite{Matthee:2024}; 6=\cite{Bogdan:2024}; 7=\cite{Napolitano:2024}; 8=\cite{Greene:2024}; 9=this paper; 10=\cite{Li:2025} }
    \label{XWebb}
\end{table}

\end{document}